%% file: carrier.tex
\documentclass[mypaper,8pt,twoside]{CoAst}
\usepackage{epsf,graphicx,fancyhdr}
\input{CoAst_liege-proceedingslayo}

\def\rfr{\smallskip\par\noindent
        \hangindent=7truemm
        \hangafter=1}

\begin{document}
\sf

\chapterCoAst{Photometric campaign on massive stars in the open cluster
\ngc{5617}}
{F.\,Carrier, S.\,Saesen, M.\,Cherix, et al.} 
\Authors{F.\,Carrier$^{1}$, S.\,Saesen$^{1}$, M.\,Cherix$^{2}$,
G.\,Bourban$^{2}$, G.\,Burki$^{2}$, J.\,Debosscher$^{1}$, D.\,Debruyne$^{1}$,
P.\,Gruyters$^{1}$, L.\,M.\,Sarro$^{3}$, M.\,Spano$^{2}$, and L.\,Weber$^{2}$} 
\Address{
$^1$ Instituut voor Sterrenkunde, Katholieke Universiteit Leuven, Belgium\\
$^2$ Observatoire de Gen\`eve, Universit\'e de Gen\`eve, Switzerland\\
$^3$ Dpt. De Inteligencia Artificial, UNED, Madrid, Spain
}

\noindent
\begin{abstract}
A campaign on the open cluster \ngc{5617} was organized in order to
characterize the pulsations and to better understand the internal structure of
its stars. The variability of the cluster members was never studied before. We
present the observations taken and an up-to-date analysis of the obtained time
series, especially of several SPB candidates we discovered.
\end{abstract}

\Session{\poster} \\ 
\Objects{\ngc{5617}} 

\section{Introduction}
Since 2003, we are monitoring a set of selected southern open clusters with the
aim of detecting and characterizing their variable stars. This program is
conducted at Euler, the 1.2-m Swiss telescope at La Silla Observatory,
Chile,
with the C2 CCD camera. The large time span of the observations, typically a
few years, makes it possible to study long-period variables, while repeated
observations during a given run allow us to detect short-term variations as
well.

The majority of the clusters is still in the process of being observed, and the obtained data
are being reduced
and
analyzed. In these proceedings we present the first results of this survey by
showing light curves of massive stars, like SPBs, but also $\delta$\,Scuti and
$\gamma$\,Dor stars, as well as eclipsing or ellipsoidal binaries.

\section{Observations and reductions}
Over a period of 5.5~years, we took 4400~measurements in the V-band and 750 in
each of the Geneva U- and B-bands. The fluxes were extracted using a
personal revised version of {\sc daophot} (Stetson~1987) using combined PSF
and aperture photometry. Then the effects of airmass, atmospheric extinction,~
\dots, were corrected using multi-differential photometry. The data
presented here
are the V-band observations and are still not detrended. Their accuracy currently reaches
2.5~mmag for the brightest non-variable stars, over a period of more than
5~years. We plan to use
the software SysRem (Tamuz et al.~2005) to correct the remaining systematic
effects. 

\section{Variable stars in \ngc{5617}}
We found a total of 218~stars displaying variability in the V-band with periods
shorter than 50~days. The stars with longer periods have to be taken with caution
(possible instrumental drifts) and will be analyzed later. Our automated
classification software (Debosscher et al.~2007) found amongst others about~35
SPB, 30~$\delta$\,Scuti and 20~$\gamma$\,Dor candidates, and 40~eclipsing and
15~ellipsoidal binaries, but the method does not take spectral information
into account (known thanks to the membership to the cluster). The variability of
these stars is thus doubtless but the classification has to be verified.
Hereafter we present some of the candidate pulsators.

\subsection{Two SPB candidates}
In the top panels of Fig.~\ref{fig}, we show the amplitude spectra in
different stages of prewhitening of two SPB star candidates. For one of the
candidates, seven significant frequencies, with values
between 0.495\,\cd and 0.784\,\cd, were detected. Peaks near 0.58\,\cd could
belong to a quintuplet $\ell=2$. The amplitudes range from 16.3 to 4.0~mmag.
For the other candidate SPB pulsator, three modes were found: $f_1$=0.677\,\cd,
$f_2$=0.574\,\cd, $f_3$=0.768\,\cd. The amplitudes are 6.3, 3.6 and 2.7~mmag
respectively.

\subsection{A $\delta$\,Scuti and $\gamma$\,Dor candidate}
In the middle panels of Fig.~\ref{fig}, we plot the amplitude spectra in
different stages of prewhitening of a $\delta$\,Scuti and a $\gamma$\,Dor
candidate. We determined seven significant frequencies in this
$\delta$\,Scuti candidate, all between 17.995 and 35.228\,\cd. The amplitudes
vary from 3.2 to 0.8~mmag. For the $\gamma$\,Dor candidate we
could identify three frequencies: $f_1$=1.746\,\cd, $f_2$=1.912\,\cd,
$f_3$=1.710\,\cd. Their amplitudes measure 14.0, 5.1 and 3.3~mmag.

\subsection{Some eclipsing binaries}
Several eclipsing binaries were identified in \ngc{5617}. In the lower panels of
Fig.~\ref{fig}, we show the phase plots for two of them: one with a short period
of P=0.296d and another one with a longer period of P=2.178d.

\figureCoAst{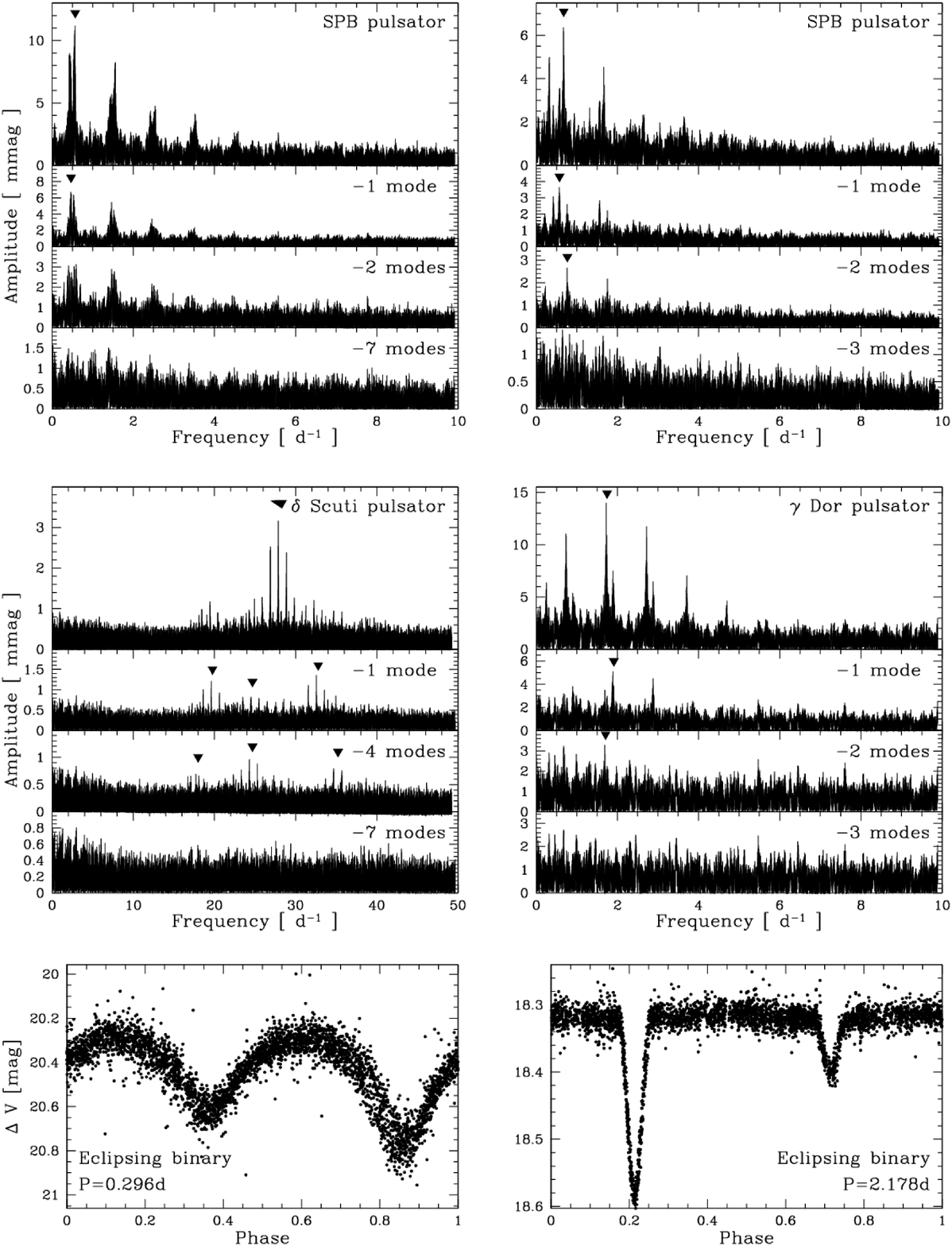}{Some example
pulsator candidates and eclipsing binaries in
\ngc{5617}.}{fig}{!ht}{clip,angle=0,width=\textwidth }

\section{Future work}
The first variability results in \ngc{5617} encourage a more in depth analysis
of the observations. We will search in detail for oscillating stars and
try to identify their main modes by means of the multi-colour photometry. With
the absolute photometry we have at our disposal, we will select some
interesting oscillating cluster members and finally attempt to model them.

\acknowledgments{
F.\,Carrier is a Postdoctoral Fellow and S.\,Saesen is an Aspirant Fellow of the
Fund for Scientific Research, Flanders (FWO). Part of this work was supported
financially by the Swiss National Science Foundation.
}

\References{
\rfr Debosscher,\,J., Sarro,\,L.\,M., Aerts,\,C. et al. 2007, \aap, 475,1159
\rfr Stetson,\,P.\,B. 1987, PASP, 99, 191
\rfr Tamuz,\,O., Mazeh,\,T., \& Zucker,\,S. 2005, \mnras, 356, 1466
}

\end{document}

%% file: CoAst_liege-proceedingslayo.tex
\renewcommand{\chaptermark}[1]{\markboth{#1}{}}

\newcommand{\aap}{A\&A}  
\newcommand{\mnras}{MNRAS}

\newcommand{\ngc}[1]{NGC~{#1}}                              

\def\cd {d$^{-1}$}

\newcommand{\kopf}{\small\itshape Comm. in Asteroseismology \\ Contribution to the Proceedings of the 38$^{th}$\,LIAC\,/\,HELAS-ESTA\,/\,BAG, 2008
}

\newcommand{\Authors}[1]{\begin{center}\normalsize\bf\sf #1 \end{center}}

\renewcommand{\author}[1]{\begin{center}\normalsize\bf\sf #1 \end{center}}
\newcommand{\Address}[1]{\begin{center}\small\sf #1 \end{center}}

\DeclareGraphicsExtensions{.eps,.jpg}

\newcommand{\Session}[1]{{\vspace{3mm}\small \noindent  \hspace*{3mm} Session: } #1 \normalsize}

\newcommand{\Objects}[1]{{\vspace{0mm}\small \noindent  \hspace*{3mm} Individual Objects: } \small #1 \normalsize}

	\newcommand{\poster}{\small Poster}

\renewenvironment{abstract}{\section*{Abstract}\normalsize\sf}{}
\newcommand{\References}[1]{\begin{flushleft}{\large References\\}\vspace*{2mm}\small #1 \end{flushleft}}

\newcommand{\chapterCoAst}[2]{\chapter[\sf\normalsize #1\\ \footnotesize \hspace*{5mm}by #2 \sf\normalsize][]{#1\\}\rhead[\fancyplain{}{\sf\footnotesize \center{#1}}]{\fancyplain{}{\sffamily\thepage}}\lhead[\fancyplain{\kopf}{\sffamily\thepage}]{\fancyplain{\kopf}{\sf\footnotesize \center{#2}}}}

\newcommand{\figureCoAst}[5]{\begin{figure}[#4]
\centering
\includegraphics*[#5]{#1}
\caption{#2}
\label{#3}
\end{figure}}

\renewcommand{\chaptername}{}

\newcommand{\acknowledgments}[1]{\vspace*{5mm}\noindent  \textbf{Acknowledgments.} #1}